# Digital Iran Reloaded: Gamer Mitigation Tactics of IRI Information Controls

Melinda Cohoon, PhD Interdisciplinary Near and Middle Eastern Studies
Information School, University of Washington
Seattle, WA, United States
mecohoon@uw.edu

*Preprint. Supported by the Open Technology Fund (OTF) Information Controls Fellowship Program (ICFP).*

Contents

## Summary Findings

## 1. Introduction

Digital authoritarianism is more rampant than ever. Authoritarian regimes use repressive information controls as a method of warfare by weaponizing surveillance and censorship, creating online deception, and a gamut of other tactics that foster information disorder in and across state borders.

Iranian Internet controls provide an interesting case study into digital authoritarianism because of the government's systematic approach to suppressing citizens online. The Islamic Republic of Iran (IRI) accomplishes this through sophisticated institutions, from the Cyber Police to the National Information Network (NIN) under the Supreme Council of Cyberspace. Through NIN, authorities have rolled out a "layered" or tiered internet structure that allows varied access to the internet, a mechanism that segregates users based on their social or political status.[1] As a deliberate strategy to suppress public and democratic discourse, managing bandwidth and access consequently impacts the human right to internet access under Article 19 of the UDHR.

Stratifying the online environment ensures that access to information and online services is unequal and, at the same time, reinforces the regime's power by coopting the internet's potential for citizen empowerment and connection. In other words, what is deemed trustworthy by the regime to enjoy freely versus what citizens might face can be a vast chasm of online experiences. This study aims to understand how Iranians counter barriers like high latency, unstable connections, and denial of access to critical online resources, especially because Iranian institutions create an artificial digital divide that stifles innovation, economic growth, and citizen engagement with the global community. Iran's use of digital authoritarianism highlights the growing trend of authoritarian regimes wielding technology to maintain power and suppress opposition.

The Digital Iran project seeks to address digital authoritarianism from the perspective of gamers, a microcosm of users in the IRI, to highlight circumvention strategies and internet control consequences. The project focuses specifically on latency, VPN usage, and gamers' perspectives as metrics to better assist high-risk internet users and to provide insights to the circumvention tools developer community.

This project was part of the Information Controls Fellowship Program (ICFP) of the Open Tech Fund (OTF), advised by Amir Rashidi from the Miaan Group.

**1.1 Context and Significance**

On September 13, 2022, Iran's Morality Police violently arrested Mahsa Amini, a 22-year-old woman of Iran's Kurdish repressed minority. She was stopped because her hijab supposedly did not meet proper standards. Hours later, she fell into a coma and then died three days later. Authorities denied any responsibility. Her death sparked a wave of protests led by women demanding justice and freedom. These protests soon swelled into larger anti-government demonstrations. In response, Iranian authorities implemented their usual crackdown tactics, including Internet disruptions, censorship, and the use of force.[2]

---

[1] Cohoon, M. (2022). Information controls in Iranian cyberspace: a soft war strategy. Doha Institute. https://www.dohainstitute.org/en/Lists/ACRPS-PDFDocumentLibrary/information-controls-in-iranian-cyberspace-a-soft-war-strategy.pdf

[2] Amnesty International. (2022, September 23). Iran: Deadly crackdown on protests against Mahsa Amini's death in custody needs urgent global action. https://www.amnesty.org/en/latest/news/2022/09/iran-deadly-crackdown-on-protests-against-mahsa-aminis-death-in-custody-needs-urgent-global-action/

As Internet shutdowns and blocking ensued, including the targeting of Microsoft products, DOTA 2, and Clash of Clans, it also appeared to discriminate against users through a tiering system, which blocked them from the global Internet and local data centers, except local businesses[3]. In the background during the Woman Life Freedom movement, there were several grumblings of an Internet User Protection Bill (IUPB) and the Seventh Development Plan to augment Internet access further.[4]

Today, the government's proposed IUPB reveals the deepening reach of digital control in Iran. Despite not having parliamentary approval, parts of this bill are already reshaping Iran's Internet landscape. At its core, the bill hands unprecedented power to the Supreme Regulatory Commission, operating under the Supreme Council of Cyberspace, to control who gets what kind of internet access and how much bandwidth they can use.[5]

What makes this particularly concerning is how the bill could affect Iran's most vulnerable communities. LGBTQ+ Iranians, for instance, rely heavily on the global internet as a lifeline to information, support, and community[6] – resources often unavailable or dangerous to access within Iran. Because the bill allows authorities to restrict access to the international internet, it threatens to cut off these vital digital connections. Even without full approval, the bill's partial implementation signals a troubling eagerness to expand digital authoritarianism under the guise of "user protection."

The pressing situation in Iran drives this project's exploration of internet restrictions from the everyday user's point of view. The state is implementing surveillance and pioneering new methods of state control over online spaces. Digital Iran is therefore concerned with these new plans and methodologies, especially as a tiered internet system impacts users' ability to access based on their status, where they work, and how much of a threat the government thinks they pose.

These controls significantly impact Iranians' day-to-day lives, even extending far beyond Iran's borders. For the Iranian diaspora, these restrictions create digital barriers between them and their homeland, affecting everything from family connections to business relationships.[7] Inside Iran, citizens face an increasingly stratified digital society, where their access to the global Internet depends on their position within the government's tiering system.

---

[3] Rashidi, A. (2022, December 24). #Women, Life, and Internet Shutdowns: Network Monitor, September 2022. Filterwatch. https://filter.watch/english/2022/10/17/women-life-and-internet-shutdowns-network-monitor-september-2022/
[4] Freedom House. (2024). Iran. In Freedom House. https://freedomhouse.org/country/iran/freedom-net/2024
[5] Article 19 (2022, September 09). Iran: Cyberspace authorities 'silently' usher in draconian internet bill. https://www.article19.org/resources/iran-draconian-internet-bill/
[6] Office of the High Commissioner (2025, March 14). Iran Government Continues Systematic Repression and Escalates Surveillance. https://www.ohchr.org/en/press-releases/2025/03/iran-government-continues-systematic-repression-and-escalates-surveillance
[7] Michaelsen, M. (2020). Silencing Across Borders: Transnational Repression and Digital Threats Against Exiled Activists from Egypt, Syria, and Iran. In Ura Design & Humanist Organization for Social Change (Eds.), HIVOS.

Despite the significance of these developments, there are substantial gaps in our understanding. While there is considerable research on internet shutdowns and conventional censorship, forms of resistance to that control – like Iranian gamers – remain understudied.

Most existing literature focuses on traditional blocking and filtering methods, leaving newer, more subtle forms of digital control underexamined. This gap is particularly problematic because these newer methods could represent the future of digital authoritarianism – less visible but potentially more effective at controlling online behavior.

Therefore, comprehensive documentation and analysis are critical. Without a detailed examination of these evolving control mechanisms, we risk missing crucial developments in how authoritarian states adapt to and co-opt digital technologies. Understanding Iran's system isn't just about documenting current practices and the future of digital control and its implications for global internet freedom.

**1.2 Research Objectives**

Digital Iran focuses on gamers as a framework for studying internet access in the Islamic Republic of Iran (IRI). This led to a comprehensive analysis of commonly used VPNs, IRI law, and latency. Gamers use VPNs to cloak their location to help circumvent Iran's censorship system, like internet tiering, bandwidth throttling, and internet filtering. These gamers, like scholars, are among the first groups to be targeted by the IRI during internet censorship crackdowns. As mentioned, the IRI explicitly blocked the online PC game DOTA 2 in late September 2022 to prevent communication during the protests. Other online games, like XBOX games and Clash of Clans, were also blocked to mitigate communication. Not only were games blocked, but so were platforms like Discord, which are essential for gamers to communicate with others in their community. While protests galvanized the IRI to restrict internet access, from throttling to shutdowns, the IRI's modus operandi has been and continues to be internet tiering. The IRI's "Seventh Development Plan" reduced international bandwidth. It encouraged users to use local services with its fiber optics rollout as part of the censorship system and tiering rollout.

The primary rationale for the IRI's implementation of the Seventh Development Plan and parts of IUPB is, by and large, a public relations strategy to control and prevent Western information from entering Iran. Indeed, there are approximately 500,000 trained workers for cyberspace for

---

https://www.opentech.fund/wp-content/uploads/2023/11/SILENCING-ACROSS-BORDERS-Marcus-Michaelsen-Hivos-Report.pdf

this program alone.[8] Consequently, several institutions implement censorship strategies, shutdowns, and other tactics. These institutions often have similar roles, suggesting a high level of government oversight. Often, the right-hand does not seem to know what the left hand is doing and vice versa, suggesting dispersed and overlapping information control tactics even if the system itself is sophisticated.

By examining gamers' use of VPNs, the Digital Iran project's key objectives are as follows:

- **Assess Internet Access:** one aim is to examine how the right to internet access is affected in Iran by focusing on gaming communities. This population is a valuable lens for understanding broader trends because of their technical sophistication and reliance on reliable, unrestricted connectivity.
- **Explore VPNs and Tiering Strategies:** another project goal is to investigate the relationship gamer experiences using VPNs against the IRI's censorship strategies. Therefore, the project examines how these tools bypass censorship and assess perceived privacy, risk, and trust.
- **Gather Technical Metrics:** A related aim is to collect and analyze concrete data such as ISP performance, bandwidth availability, ping times, packet loss, and other related metrics. This helps determine how effectively censorship and surveillance measures work and the disruption they cause users.

## 2. Research Methodology and Design

This study employs a mixed-methods approach to examine internet access and digital censorship in Iran, using gaming communities as a strategic lens. The research was conducted over a one-year period and progressed through three distinct phases.

Phase one consisted of an initial qualitative research phase of approximately five months. This involved in-depth interviews with 12 participants from Iranian gaming communities. These preliminary interviews served a dual purpose: (1) establishing further rapport with the gaming community and (2) informing the development of survey questions for the subsequent quantitative phase. After identifying key challenges and adaptive strategies employed by gamers, I began the second phase of the research.

Phase two consisted of interviews and survey distribution of approximately four months. Following the initial interviews, the survey was disseminated to a broader audience. Concurrently, additional interviews were conducted, bringing the total number of interview respondents to 22. This phase integrated qualitative data from extended interviews, quantitative data from survey responses, and ethnographic observations of participant behaviors and community dynamics.

---

[8] Farda-ye Eghtesad. (2023, October 24). The government is training 500,0000 people for cyberspace +video https://www.fardayeeghtesad.com/news/31420/%D8%AF%D9%88%D9%84%D8%AA-%DB%B5%DB%B0%DB%B0-%D9%87%D8%B2%D8%A7%D8%B1-%D9%86%DB%8C%D8%B1%D9%88-%D8%A8%D8%B1%D8%A7%DB%8C-%D9%81%D8%B6%D8%A7%DB%8C-%D9%85%D8%AC%D8%A7%D8%B2%DB%8C-%D8%AA%D8%B1%D8%A8%DB%8C%D8%AA-%D9%85%DB%8C-%DA%A9%D9%86%D8%AF-%D9%81%DB%8C%D9%84%D9%85.

Phase three consisted of technical analysis and synthesis of the research for the remainder of the fellowship period. The final phase incorporated network analysis to measure latency and other technical parameters affecting gaming experiences. This technical data correlated with the qualitative findings to provide a comprehensive understanding of digital censorship's practical impacts, users' adaptive strategies, and patterns of socio-political trust within gaming communities.

Altogether, the research methodology and design had significant implications, which are further elaborated in the findings section of the report. Gaming communities provided a lens into an ideal population for this research because they regularly navigate censorship barriers to access international gaming platforms. Additionally, their proficiency with VPNs and other circumvention tools provides insights into practical workarounds. Their experiences also illustrate the broader implications of digital control measures on everyday internet users. Gaming communities foster social connections that reveal patterns of trust and information sharing.

**2.1 Data Collection**

*2.1.1 Qualitative Data*

In preparation for this report, interviews were conducted with 22 respondents who are self-described gamers in Iran. Respondents included 33% women and 66% men, aged 24 to 35, who live in Tehran. While these narratives collected from the 22 respondents represent a small ecosystem of gamers in Iran, the report acknowledges the time constraints and the overall intricate details that interviews provide, illuminating the everyday life experiences of Iranians. Most interviewees would certainly not consider their actions to attain access as subversive or inherently political, as most Internet users in Iran must use a VPN at one time or another. So, interviewees would agree that Internet access is an inherent human right, and gaming is just one of life's pleasures that a filtered Internet should not impact. One way to describe gamers' actions and sentiments could be sociotechnical resilience to surveillance and censorship. Iranian gamers often rely on a combination of state-of-the-art circumvention tools and grassroots practices to bypass digital censorship.[9] Sociotechnical resilience[10] emphasizes how these individuals build, test, and refine their adaptive strategies in response to technical restrictions (e.g., blocked websites, patched VPN vulnerabilities) and evolving surveillance tactics.[11]

---

[9] Dehshiri, A. (2024, May 17). #The use of VPNs is prohibited, but not criminalized. Filterwatch. https://filter.watch/english/2024/03/04/the-use-of-vpns-is-prohibited-but-not-criminalized/

[10] Golub, A. (2010). Being in the World (of Warcraft): raiding, realism, and knowledge production in a massively multiplayer online game. *Anthropological Quarterly*, *83*(1), 17–45. https://doi.org/10.1353/anq.0.0110

Respondents were interviewed during the first two phases of the research. Initial interviews were conducted from February 2024 through June 2024 through a preexisting network of individuals based on established contacts from the researcher during the dissertation research conducted for *Affective Lifeworlds: Iranian Gamers vs. the Islamic Republic of Iran*.[12] During the initial and second phases, recruitment through snowball sampling also occurred, which led to a second round of interviews during the second phase of the research. By building on this preexisting network, a secure network of trust was built, leading to a total of 22 respondents. Not all contacts invited for an interview via this network ultimately agreed to participate. Final interviews were conducted in September 2024.

Using semi-structured guidelines, the interviews proceeded with room for conversation outside of the scripted questions. Respondents touched on their favorite games and VPNs, challenges they have had accessing the Internet, their ability or inability to play certain games, perspectives on VPN trustworthiness, and experiences of threats or fear of surveillance. These interviews were transcribed, translated into English, and coded using ATLAS.Ti qualitative analysis software. Interviews were all conducted through Signal.

In addition to interviews, the research also included participant observation of gamers online through their private Discord servers. Participant observation of Discord gaming communities reveals how Iranian players develop sophisticated sociotechnical resilience strategies, simultaneously navigating censorship barriers through technical workarounds while cultivating trusted networks where knowledge about VPN configurations, server access, and security practices is collectively shared and refined through daily interactions. These participant observation notes were transcribed, translated into English, and coded using ATLAS.Ti qualitative analysis software. The coded material from interviews and participant observation led to general themes that assisted in developing the survey questions.

*2.1.2 Quantitative Data*

During the second phase, the study's survey was distributed to a broader audience of Iranian gamers, capturing quantitative metrics about their digital experiences under censorship. The survey participants amounted to 666 responses, with some cut out for trolling, and therefore narrowed slightly to 660 relevant responses. The survey had 16 questions, including demographics, trust in VPNs, and the political trust of IRI Internet access policies like the Seventh Development Plan.

---
[11] Egherman, T. (2025, March 19). *Iran's Digital Control: the evolution of censorship and surveillance amidst the 'Women, Life, Freedom' movement*. Miaan Group. https://miaan.org/irans-digital-control-the-evolution-of-censorship-and-surveillance-amidst-the-women-life-freedom-movement/
[12] Cohoon, M. (2022, December 1). Presented at the Middle East Studies Association Conference. https://melindacohoon.com/presented-at-the-middle-east-studies-association-conference/

Evaluating survey responses necessitated reading individual responses closely and analyzing them using ATLAS.Ti, in addition to using other tools to generate visual models, i.e., Excel for pie charts, Python libraries like Matplotlib for heat maps, etc.[13] The closed-ended questions particularly necessitated hierarchical regression analysis to provide a deeper understanding of political trust. In other words, hierarchical regression analysis was essential due to the layered structure of the closed-ended questions, which captured multiple levels of influence on political trust. Sequentially introducing predictors into the model allowed the study to first account for baseline socio-demographic factors and then assess the incremental impact of political attitudes and context-specific variables. This stepwise dissection revealed the direct effects of key predictors and how their relationships with political trust evolved when controlling other confounding factors, ultimately leading to a more nuanced understanding of the complex determinants of political trust.

Phase three consisted of network analysis and synthesis of all data, which included analyzing Round-Trip Time (RTT) to evaluate the practical impacts of digital restrictions for gamers. In gaming, RTT is essentially what gamers experience as ping or latency, measured in milliseconds (ms). Lower RTT values indicate better network performance. This is important for real-time applications and platforms like online gaming, where high RTT can cause lag and affect gameplay. Most gamers find 100 ms or higher unacceptable or unplayable.[14] However, 100 ms or higher is often the experience for Iranian gamers without a VPN.[15] So, to play games effectively, Iranian gamers need lower ping, often necessitating better VPNs and server access outside of Iran. RTT measurements tell us something important about the performance of using VPNs or other circumvention tools, helping identify which servers or regions give gamers the most responsive connections and providing objective evidence of how censorship affects user experience.[16]

## 2.2 Analytical and Methodological Framework

Through the research design, Digital Iran integrates the following analyses:
- Technical metrics concerning internet performance and security risks
- User experiences and adaptation strategies in navigating censorship

---

[13] Stewart, L. (2025, February 11). *Best Data Visualization Tools | Comparison & Overview*. ATLAS.ti. https://atlasti.com/research-hub/data-visualization-tools

[14] G. Martínez, J. A. Hernández, P. Reviriego and P. Reinheimer. (2024). Round Trip Time (RTT) Delay in the Internet: Analysis and Trends. *IEEE Network 38*(2), 280-285. doi: 10.1109/MNET004.2300008.

[15] Financial Tribune. (2019, February). High Ping, Internet Disruptions Make Online Gaming Hard in Iran. https://financialtribune.com/articles/sci-tech/96737/high-ping-internet-disruptions-make-online-gaming-hard-in-iran

[16] Al Jazeera (2019, December 23). Locked Out: US sanctions are ruining online gaming in Iran https://www.aljazeera.com/economy/2019/12/23/locked-out-us-sanctions-are-ruining-online-gaming-in-iranl j

- Assessment of how these factors influence domestic internet accessibility and connectivity.

Digital Iran's methodological approach is fundamentally grounded in sociotechnical resilience, a framework that examines how Iranian internet users, particularly gamers, develop adaptive capacities at the intersection of technical systems and social practices. This lens recognizes that resilience against digital censorship emerges from technological solutions and the complex interplay between technical workarounds and social support networks.

The multi-layered analysis captures how resilience manifests across different dimensions of the Iranian digital ecosystem. By measuring technical metrics like RTT (Round-Trip Time), packet loss, and VPN performance alongside qualitative data on user experiences, the research reveals the dynamic processes through which gamers maintain connectivity despite increasingly sophisticated censorship measures.

The quantitative component of Digital Iran focuses on measurable indicators of network performance under censorship conditions using RIPE Atlas.[17] These metrics serve as tangible evidence of both state restrictions and user adaptations. For instance, comparing ping times with and without circumvention tools quantifies gamers' technical trade-offs when prioritizing access over performance. Similarly, patterns in packet loss during peak censorship periods illuminate how the technical infrastructure responds to state intervention.[18]

These metrics go beyond simply documenting censorship effects because they reveal the technical dimension of resilience by showing how systems adapt, where breaking points occur, and which workarounds prove most effective. The hierarchical regression analysis applied to these metrics helps identify which technical factors most significantly influence users' ability to maintain functional connections to international gaming servers. In applying hierarchical regression analysis, this study sought to elucidate sociotechnical and political trust. Studies of gamers have shown that they are:

- More likely to encounter trolling and other antisocial behaviors in online gaming environments, which may negatively shape their views on society.
- Experience a sense of self-worth and social status within the game world that does not translate to the real world, leading them to disengage from sociopolitical life.

Essentially, because gaming is becoming an increasingly popular pastime, there is potential for gamers to disconnect from sociopolitical life and develop less prosocial attitudes.[19] Hierarchical regression analysis affords more nuance and complexity to this dilemma, showing that while most Iranian gamers see gaming as non-political, that state's intrusion impacts decision-making and trust, ultimately affording more socio-political habits related to gaming as a pastime. In other words, though gaming itself may not be inherently political for many Iranian gamers, state intervention and censorship effectively politicize the activity. Iranian authorities impose information controls directly impacting gamers' access to international servers and gaming platforms. These restrictions force gamers to adopt sophisticated circumvention techniques (VPNs, DNS services, tunneling methods, etc.) to participate in what would otherwise be a non-political leisure activity.

---

[17] RIPE Atlas. (n.d.). About RIPE Atlas. https://www.ripe.net/analyse/internet-measurements/ripe-atlas/#:~:text=RIPE%20Atlas%20is%20a%20global,data%20visualisations%2C%20and%20an%20API.
[18] Payande, I. (2024, July 9). Censorship and sanctions impacting Iran's internet, report. Internet Society Pulse. https://pulse.internetsociety.org/blog/censorship-and-sanctions-impacting-irans-internet-report
[19] Bacovsky, P. (2020). Gaming alone: Videogaming and sociopolitical attitudes. *New Media & Society*, *23*(5), 1133–1156. https://doi.org/10.1177/1461444820910418

Although gamers often believe that they are socio-politically disengaged,[20] Iranian gamers demonstrate enhanced prosocial behaviors through necessity that are less evident in Western contexts.[21] The state's intrusion into gaming spaces[22] has inadvertently fostered community-oriented practices where gamers actively share technical knowledge, collaborate on circumvention strategies, and build resilient social networks.

Rather than a cyber utopia, Iranians experience a weaponization of the Internet and other digital media by the regime's Internet control bureaucracy–from the Islamic Revolutionary Guard Corps (IRGC) to the newly minted Department of Security, Integration, Innovation that recently sanctioned the Seventh Development Program by the government.[23] The regime's digital control apparatus continues to expand, as evidenced by the Supreme Council of Cyberspace's Resolution No. 3 on March 1, 2024, which banned VPN usage for all users.[24] This resolution exemplifies the contradictory nature of Iran's digital governance. While implementing a comprehensive VPN ban, the Ministry of ICT is simultaneously tasked with providing "sanction-breaking services" through licensed private sector entities.[25] These intensifying restrictions, documented through blocking over 900 VPN services and increased surveillance of Internet cafes, demonstrate the regime's systematic approach to digital control.[26]

---

[20] Losh, E. (2016). Hiding Inside the Magic Circle: Gamergate and The End of Safe Space. b2o: boundary 2 online. https://www.boundary2.org/2016/08/elizabeth-losh-hiding-inside-the-magic-circle-gamergate-and-the-end-of-safe-space/
[21] Cohoon, M. (2022). Digital Borderlands: Soft War as Discourse in Iranian Video Games. *IDEAH, 3*(2). https://doi.org/10.21428/f1f23564.73008090
[22] Corera, G. (2021, February 8). Iran 'hides spyware in wallpaper, restaurant and game apps." BBC. https://www.bbc.com/news/technology-55977537
[23] Farda-ye Eghtesad. (2023, October 24). The government is training 500,0000 people for cyberspace +video https://www.fardayeeghtesad.com/news/31420/%D8%AF%D9%88%D9%84%D8%AA-%DB%B5%DB%B0%DB%B0-%D9%87%D8%B2%D8%A7%D8%B1-%D9%86%DB%8C%D8%B1%D9%88-%D8%A8%D8%B1%D8%A7%DB%8C-%D9%81%D8%B6%D8%A7%DB%8C-%D9%85%D8%AC%D8%A7%D8%B2%DB%8C-%D8%AA%D8%B1%D8%A8%DB%8C%D8%AA-%D9%85%DB%8C-%DA%A9%D9%86%D8%AF-%D9%81%DB%8C%D9%84%D9%85.
[24] Radio Farda (2024, February 20). The use of filter breakers in Iran was officially banned after approval from the Leader of the Islamic Republic. https://www.radiofarda.com/amp/iran-announce-vpn-illegal-Internet-khamenei/32827817.html
[25] Article 19. (2024, July 23). Tightening the Net: Iran's New phase of digital repression. https://www.article19.org/resources/tightening-the-net-irans-new-phase-of-digital-repression/.
[26] Dadbazar. (2024, February 1) Resolution of the Supreme Cyberspace Council on Combating Filter Violators. https://www.ekhtebar.ir/%d9%85%d8%b5%d9%88%d8%a8%d9%87-%d8%b4%d9%88%d8%b1%d8%a7%db%8c-%d8%b9%d8%a7%d9%84%db%8c-

By looking at VPNs commonly used by gamers using qualitative methods, this study shows how Iranian gamers collectively develop, share, and refine adaptation strategies. Through participant observation in Discord communities and in-depth interviews, the research documents the social processes that enable knowledge transfer about effective circumvention techniques. These communities function as resilience hubs where technical expertise is democratized and distributed across networks of trust.

Iranian gamers and online activists carve out online spaces despite facing systemic efforts to codify and constrain their everyday digital interactions—efforts aimed at steering narratives to serve ulterior political purposes.[27] To do so, they must counter systematic regulation, such as Iran's ongoing development of its National Information Network infrastructure, intensified Internet restrictions during periods of social unrest, and expanding digital surveillance measures.[28] While facing systemic efforts to codify and constrain their everyday digital interactions—efforts aimed at steering narratives to serve ulterior political purposes—Iranian gamers and online activists carve out online spaces. Gamers in Iran keep finding new ways to hang out online. Some build their chat rooms in Discord, and others meet up in-game worlds.

This research focus lies in integrating these technical and social dimensions. By combining quantitative metrics with qualitative insights, the research illuminates how sociotechnical resilience emerges as a property of the entire system rather than individual components. For example, the study reveals how periods of intensified technical restrictions often coincide with heightened community collaboration and knowledge sharing.

The hierarchical regression analysis further enriches this integrated understanding by controlling baseline factors before examining how technical capabilities, social connections, and political attitudes collectively influence users' resilience strategies. This approach reveals what technical solutions work and how social factors mediate their effectiveness and adoption.

---

%d9%81%d8%b6%d8%a7%db%8c-%d9%85%d8%ac%d8%a7%d8%b2%db%8c-%d8%af%d8%b1%d8%ae%d8%b5%d9%88%d8%b5-%d9%85%d9%82%d8%a7/. ISNA (2023, March 26). A look at the process of completing the national information network. https://www.isna.ir/news/1402010601938/%D9%86%DA%AF%D8%A7%D9%87%DB%8C-%D8%A8%D9%87-%D8%B1%D9%88%D9%86%D8%AF-%D8%AA%DA%A9%D9%85%DB%8C%D9%84-%D8%B4%D8%A8%DA%A9%D9%87-%D9%85%D9%84%DB%8C-%D8%A7%D8%B7%D9%84%D8%A7%D8%B9%D8%A7%D8%AA

[27] Keshavarznia, N. (2024, September). The Monopoly of Internet Infrastructure in Iran: Challenges and Consequences of Restrictions on Access and Cybersecurity. FilterWatch. https://filter.watch/english/2024/10/10/https-filteinvestigative-report-september-2024-Internet-infrastructure-monopoly/. Aryan, S., Aryan, H., & Halderman, J.A. (2013). Internet Censorship in Iran: A First Look. *FOCI 13*. https://www.usenix.org/system/files/conference/foci13/foci13-aryan.pdf

[28] Mahoozi, S. (2023, September). FEATURE-Iran steps up Internet crackdown one year after Mahsa Amini's death. Reuters. https://www.reuters.com/article/business/media-telecom/feature-iran-steps-up-Internet-crackdown-one-year-after-mahsa-amini-death-idUSL8N3AJ203/

Sociotechnical systems allow us to understand how technical elements (such as VPNs, game mechanics, and digital restrictions) interweave with social practices (community building, raid coordination, and cultural adaptation) to create complex gaming environments. This sociotechnical resilience framework ultimately provides a more nuanced understanding of how censorship impacts domestic internet users.[29] The research demonstrates that resilience is not evenly distributed. It varies according to technical literacy, social capital,[30] and resource access. These variations significantly affect who maintains connectivity during censorship events and how information flows between domestic and diaspora communities.

By interviewing gamers, Digital Iran offers insights into the evolving relationship between state control and citizen agency in digital spaces. It reveals how gaming communities, through their technical sophistication and social cohesion, often serve as critical bridges, maintaining connections across digital borders despite increasingly sophisticated censorship regimes.

## 3. Findings

### 3.3 Narratives

Gaming has become a space for many worldwide to interact with one another. However, not everyone has the same level of access from country to country. Where digital avatars traverse virtual worlds free from geographic limitations, Iranians have the same level of access as those without information controls. Among these frontline netizens are Iranian gamers who, in their pursuit of both recreation and resistance, confront a state that seeks to monitor, restrict, and control communication. Iranian gamers' experiences offer a distinctive perspective on informational controls wherein personal freedom is pitted against an omnipresent, repressive government.

Virtual worlds for Iranian gamers are more than spaces of entertainment. Iranian gamers often see online video gaming as a space to engage with their peers and an essential aspect of their lives; however, it has become another site of surveillance and censorship. Babak, a pseudonym used to protect his identity, recalled operating under this oppressive system:

***"You know what? I mostly try to ignore the government. I think that's what gaming does for me. But every time I am online, I feel a sense of dread, as if the regime's eyes are on me, even with a VPN. They can track anything."***

Mariam also chimed in:

---

[29] Golub, A. (2010). Being in the world (of Warcraft): Raiding, realism, and knowledge production in a massively multiplayer online game. *Anthropological Quarterly, 83*(1), 17-45. http://www.jstor.org/stable/20638698
[30] Zhong, Z. (2011). The Effects of Collective MMORPG Play on Gamers' Online and Offline Social Capital. *Computers in Human Behavior 27*(6), 2352–2363. doi:10.1016/j.chb.2011.07.014.

*"When the Internet is disconnected, it's like being alone in the world. But when access goes uninterrupted, it's like a very calm experience because I have my friends available just at the grasp of my hands. An Internet ban also just doesn't do very much. Regulations just incentivize us to use VPNs to have the same access as the rest of the world."*

Moreover, Mariam's characterization of uninterrupted access as "a very calm experience" reveals how virtual intimacy functions as an emotional infrastructure. Mariam's experience invokes affective dimensions of digital connectivity, where stable Internet access creates a sense of psychological security through sustained virtual presence.[31] The contrast between the "calm" of connection and the isolation of disconnection underscores how virtual intimacy has become fundamental to emotional regulation and social well-being. These constructions of identity and digital space through emotional bonds form a resilience to digital authoritarianism.

The desire for access emerged in Mariam's discussion of VPNs and global access. When she stated that "an Internet ban also just doesn't do very much," she showed how gaming becomes a site of resistance against digital authoritarianism. Virtual desire extends beyond personal relationships to encompass a broader sense of global belonging and rights. The desire to "have the same access as the rest of the world" demonstrates an understanding of digital citizenship. Deep within these online communities thus lies a network of trust, an unspoken bond forged over shared experiences of state intrusion. Members speak candidly about digital barricades and the relentless efforts of the state to stifle dissent. Arman, a seasoned gamer, lamented:

*"The government targets political voices. We want an escape or a brief, unmonitored respite in our games […] but we cannot even do something so simple."*

As authoritarian states like Iran refine their approaches to digital control, the experiences of Iranian citizens, particularly their strategies of resistance and adaptation, provide crucial insights into the ongoing struggle between state control and digital rights. The findings suggest that effective resistance to digital authoritarianism requires technical solutions, the cultivation of resilient social networks, and international solidarity.

In response, digital resistance thrives within these communities. On the technical front, Iranian gamers meticulously monitor network performance, tracking metrics like Round-Trip Time (RTT), packet loss, and VPN integrity.[32] Participants offer tangible insights into how censorship tangibly degrades the gaming experience. As one participant explained during an online discussion:

*"My ping times can suddenly spike during DOTA 2 because of deliberate throttling […] the state filters the game server, so you have to use a VPN, but the internet connection is terrible.*

---

*We cannot buy games we like with our own money because of sanctions and government dysfunction."*

Javad stated:

*"Ping is terrible in Iran. When I go onto a network outside of Iran, I still sometimes get numbers in the high 100s to 200s, sometimes even 300s! Some recommendations have been to test server connectivity on another Middle Eastern server to improve the situation. Because of bad ping, even with DNS, I cannot log onto certain games after purchasing them for PC. Sometimes, there is no tactic. But I have found Western EU servers to be better."*

These anecdotes from citizens on the ground provide evidence of users' challenges. At the same time, this user group relies on the collaborative sharing of workarounds to foster resilience, transforming individual technical struggles into collective knowledge to overcome digital obstacles.

Parallel to these technical measures, the social strategies embedded in gaming communities become equally vital. Forums, chat groups, and video game live streams are platforms where narratives of fear, resistance, and hope circulate widely. Such spaces permit the candid expression of the emotional toll of constant surveillance. Through participant observation on platforms like Discord, these communities actively disseminate information on using VPNs safely, changing configurations in response to evolving state tactics, and even disguising digital footprints to evade detection.

The participants' experience thus illustrates how the Iranian government's technical threat landscape becomes an adversary. The interrelationship between adaptation and solidarity allows Iranian gamers to reinvent their online engagement methods, turning their gaming platforms into bastions of digital resilience.

The state's surveillance apparatus, with its sophisticated tracking systems and internet throttling tactics, casts a long shadow over these interactions. Yet, it is within this shadow that resilience blossoms. Iranian gamers have cultivated an acute sense of sociotechnical resilience—a symbiosis of technical expertise and community solidarity that allows them to circumvent barriers while preserving their cultural identity and shared language of resistance.

Technical countermeasures are integral to this resilience. Gamers utilize VPNs, proxy servers, and encryption tools to obscure their digital footprints. Yet, gamers remain undeterred even as packet loss and increased latency become the norm during high-censorship periods. They must continuously optimize their settings, inquire about bypass methodologies, and share their own configurations with trusted friends. One participant in a heated conversation noted:

*"We trade our exploits like rare loot in a raid, each piece adding a small win against the enemy."*

This metaphor, rich with gaming imagery, underscores the innovative spirit and the resilience underpinning their digital resistance. Every workaround and security patch is not just a technical solution but a covert act of rebellion, reclaiming digital space that would otherwise be under the authoritarian thumb.

The impact of state repression is not limited to the technical realm, as it profoundly affects gamers' emotional and psychological well-being. Many express a familiar sense of vulnerability and isolation as trust becomes a precious commodity. Another participant, who voiced on a chat channel, confided:

***"The fear isn't just of getting caught; it's also the fear of losing our only connections. Every friend here is a lifeline."***

Such testimonials reveal the double-edged nature of online communities in Iran: they are both safe harbors and targets. The state's efforts to dismantle these networks by cutting off access and spreading misinformation have made every interaction a potential risk. Yet, paradoxically, this threat has also galvanized the community, prompting more audacious, creative forms of communication.

In parallel to these social adaptations, a crucial technical narrative unfolds. Analyses of internet performance metrics reveal that Iranian gamers continue to craft ingenious countermeasures even as the state forces routers to slow down connections and disrupt data packets. For instance, comparative studies of Round-Trip Time (RTT) and packet loss illustrate that while circumvention tools inevitably introduce lag, the gamers' careful selection of VPN servers and constant settings adjustment have minimized these performance penalties. Despite relentless throttling, their persistence in optimizing network performance is a testament to their technical resolve that reflects a broader, organized resistance.

The militarization of digital control in Iran has created a paradoxical situation where attempts to suppress digital resistance have instead fostered more sophisticated forms of circumvention and community resilience.[33] This is particularly evident in gaming communities, where networks of outrage and hope provide a lifeline across digital platforms that ultimately foster resistance and solidarity. The technical suppression mechanisms, from DNS poisoning to VPN detection, have inadvertently catalyzed the development of these small media resistance networks. These networks, operating through gaming platforms and other digital spaces, demonstrate how authoritarian Internet control forces users to live on the fringes, wherein simply playing games acts as an antidote to everyday feelings of oppression. Among those who voice this sentiment are two long-time friends, Arman and Babak, who stand out for their fearless yet subtle defiance. Their conversational blend of playful banter and coded language emphasizes a paradoxical world

---

[33] Jones, M.O. (2022). *Digital Authoritarianism in the Middle East: Deception, Disinformation and Social Media*. Oxford University Press.

where every tactical move in DOTA 2 is layered with the awareness that their online actions might become fuel for the authorities' arsenal.

*"We're winning battles on the digital front, but it always feels like the state is lurking in every server we log onto."*

*"Each strategic play is both a game move and a statement, an act of resistance in an environment that strips away our freedoms."*

This reveals how gamers' daily struggle is not solely against their adversaries in the game but against a hostile regime that monitors their connectivity and disrupts their communications. Their words carry the weight of a reality where leisure has become intertwined with survival, and each online encounter is punctuated with the unspoken threat of state reprisal.

The conversation on the server, punctuated by candid expressions of frustration and ingenuity, demonstrates that these gamers are not passive victims of censorship but active participants in digital resistance. Behind every joke and every code word lies a complex interplay of technical expertise and collective vigilance. Members share instructions on concealing IP addresses, discuss obfuscation methods, and even exchange tips on bypassing newly enforced firewalls. In these moments, the community's sociotechnical resilience comes to the fore—a dynamic process born out of a necessity to always stay one step ahead of government algorithms.

For gamers, humor functions as a strategy of resilience, transforming a moment of constraint into playful subversion. It serves as both a shield and a signal or way to diffuse the tension inherent in their shared experience while simultaneously alerting fellow gamers to subtle changes in the threat landscape. For instance, Mariam remarked:

*"The state is just there to grief us."*[34]

The state's approach to surveillance, messaging, and digital repression has drawn on age-old strategies yet reconfigured them to exert influence within cyberspace. However, camaraderie forms among gamers through technical survival tactics shaped by censorship practices. While there may be personal cost among these users, i.e., exposing themselves to potential state retaliation, gamer vigilance fosters a unique kind of community where every interaction is both an act of normalcy and an act of rebellion.

At the same time, some gamers view this as a "hidden war" on Internet access in Iran.[35] Continued efforts in Iran to tier the internet, filter, and throttle it suggest a war on Internet usage since its inception, even if the IRI makes lackluster attempts to obfuscate its goal by supposed fiber optic rollouts. The Iranian government thus employs a range of digital authoritarian tactics, using digital technologies to monitor, control, and restrict Internet access. This digital repression is part of a broader strategy of information warfare aimed at maintaining the regime's grip on power.

---

[34] Griefing means intentionally annoying another player to disrupt their gameplay. This is especially relevant in multiplayer online games. Some players do this to seek attention or have a vendetta against another gamer. They are also called bad-faith players. In the case of the Iranian state, Mariam is referring to the troll-like nature of state control. State-aligned trolling also exists as made evident in Kargar and Adrian Rauchfleisch's "State trolling in Iran and the double-edged affordances of Instagram" (2019).
[35] Shakibi, L. (2024, March 18). The hidden war on Internet access in Iran. Filterwatch. https://filter.watch/16nglish/2024/03/18/council-hacked-document-mar-2024/

This digital repression is part of a broader strategy of information warfare aimed at maintaining the regime's grip on power. No law justifies the actions of the Iranian government, but they continue to disrupt Internet access, surveil citizens, and arrest individuals for their online activities. For gamers like Babak, this "hidden war" is not just a theoretical concept but a daily reality. Living alone in a small apartment in Tehran, Babak has created a gaming setup that doubles as a site of resistance. His desk, cluttered with multiple screens and a perpetually running VPN, is a testament to the lengths he must go to access online spaces. "It's like breathing now," he said, laughing. "You don't think about it, you just do it." For Babak, gaming is not just a hobby; it is a lifeline to a world beyond the constraints of the Iranian state.

These stories emphasize the resilience and ingenuity of Iranian gamers, who navigate a labyrinth of state-imposed restrictions to maintain their connection to the digital world. In the face of threats of arrest and imprisonment, these strategies—switching between VPNs, streaming under pseudonyms, and mapping Internet disruptions—are acts of everyday resistance that challenge the state's control over digital spaces.

**3.2. Survey**

The comprehensive survey represents the methodological cornerstone of this research, offering unprecedented insights into how Iranian gamers navigate, resist, and circumvent state-imposed information controls. As the principal investigator for Digital Iran Reloaded, I designed this survey to move beyond technical assessments of censorship mechanisms toward a holistic understanding of sociotechnical resilience in practice.

With 660 responses from across Iran's urban centers, our survey captures the lived experiences of gamers developing adaptive capacities at the intersection of technical systems and social practices. The demographic diversity of our respondents—spanning various age groups, educational backgrounds, and geographic locations—has enabled us to identify patterns in circumvention strategies that would remain invisible through purely technical analysis. Particularly noteworthy is our data on gender differences in digital navigation strategies, revealing how female gamers develop distinct forms of sociotechnical resilience in response to Iran's gender-segregated digital surveillance. These findings challenge conventional approaches to digital resistance that often overlook how social positionality fundamentally shapes both the experience of censorship and the capacity for circumvention.

The survey findings in section 3.2 directly advance the project's central objective: to understand the evolving dialectic between state control and citizen agency in Iran's digital ecosystem. By integrating demographic data with a detailed analysis of gaming practices and circumvention strategies, we have documented how ordinary Iranians function not as passive victims of digital authoritarianism but as innovative agents of sociotechnical change. This approach transforms our understanding of digital resistance from a purely technical phenomenon to a complex social practice embedded within specific cultural contexts. As we continue our research with the Open Technology Fund, these survey findings provide crucial insights for multiple stakeholders—from international organizations supporting internet freedom to technology developers designing more effective and contextually appropriate circumvention tools. Most importantly, by documenting the creative resilience of Iranian gamers, our research contributes to a broader scholarly and practical understanding of how digital spaces become sites of contested authority and emergent freedom even under conditions of intensifying state control.

### 3.2.1 Demographics

The sociotechnical resilience framework offers a powerful theoretical lens for understanding how Iranian gamers navigate digital restrictions. This framework recognizes that resilience emerges through technical solutions and the complex interplay between social structures, cultural practices, and technological systems. Demographics fundamentally shape this interplay.

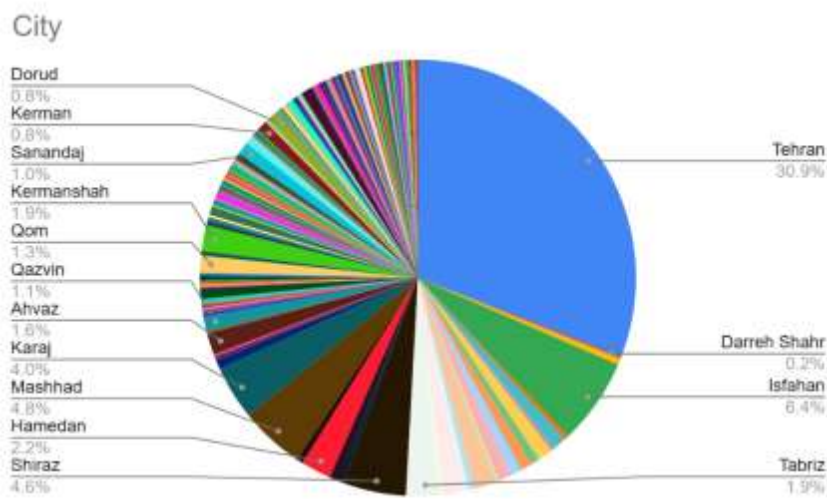

Figure 1. City distribution of 660 respondents.

The urban concentration of respondents (63% from Iran's top 15 cities) highlights the geographically distributed nature of sociotechnical resilience. Urban environments foster resilience through denser social networks that facilitate knowledge sharing about circumvention tools, creating what scholars call "communities of practice" where technical expertise is collectively developed and transmitted. These urban knowledge ecosystems enable gamers to rapidly adapt to evolving censorship mechanisms through social learning processes that are less accessible in rural settings.

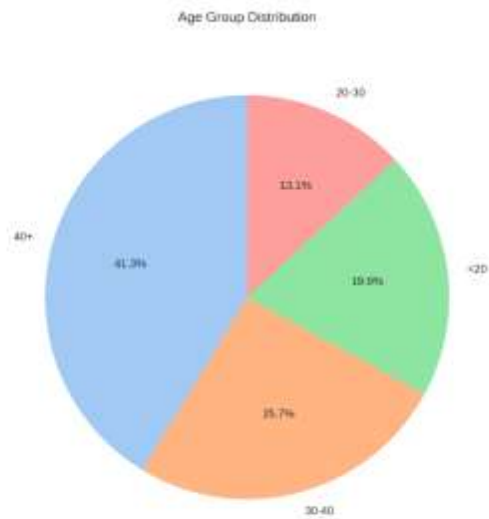

Figure 2. Age distribution pie chart of 660 respondents.

Age demographics illuminate generational variations in sociotechnical resilience strategies. The even distribution across age groups (20-30, 30-40, and 40+) allows researchers to examine how different generations integrate technical solutions into their existing social practices. Younger gamers may develop resilience through rapid technological experimentation and peer-to-peer knowledge networks, while older gamers might rely on more established social connections and institutional knowledge. These age-based variations reflect what resilience theorists call "response diversity" – how communities respond to similar challenges.

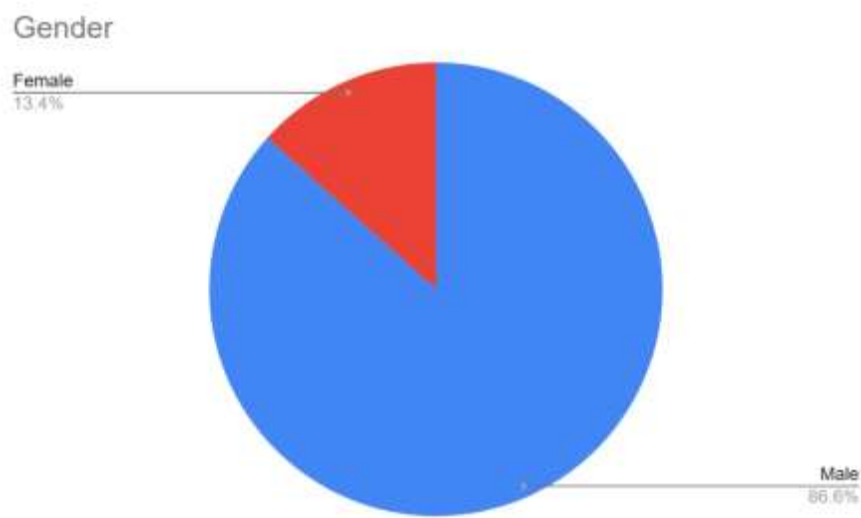

Figure 3. Gender distribution of 660 respondents.

Gender represents a particularly crucial demographic factor in understanding sociotechnical resilience among Iranian gamers, though it appears underexplored in the report. In Iran's gender-segregated society, women and men likely develop distinctly gendered forms of sociotechnical resilience.

Female gamers may face additional layers of surveillance and restriction, potentially driving the development of gender-specific circumvention strategies. Research on digital resilience in restrictive contexts suggests that women often create hidden or encrypted communication channels within seemingly apolitical platforms like gaming networks. These "hidden transcripts" of resistance allow female gamers to build resilience through gender-specific social networks that combine technical workarounds with social camouflage.

Gender also influences access to technical knowledge and resources. In patriarchal contexts, technical expertise is often gendered as masculine, potentially limiting women's access to formal training in circumvention tools. However, sociotechnical resilience theory suggests that marginalized groups usually develop alternative knowledge pathways. Female gamers may cultivate resilience through women-specific knowledge networks that adapt circumvention tools to their particular social constraints.

The intersection of gender with other demographic factors like education and urban location further complicates these dynamics. Highly educated urban women may develop different resilience strategies than their rural counterparts, highlighting what resilience theorists call "cross-scale interactions" – how broader social systems shape resilience at the individual level.

By examining these demographic dimensions through a sociotechnical resilience framework, we gain deeper insight into how Iranian gamers don't merely use circumvention tools but develop integrated sociotechnical practices that embed technical solutions within existing social structures. This theoretical grounding transforms demographic data from descriptive statistics into analytical tools for understanding how different population segments develop unique adaptive capacities at the intersection of technical systems and social practices.

### *3.2.2. Survey Analysis*

After analyzing the qualitative data, I created heatmaps for two of the open-ended questions since they specifically relate to one of the study's five major questions: **With pervasive internet restrictions, how do gamers maintain their level of access to video games?**

<u>Definitions</u>

**Generic VPN:** VPN is listed but not specified by the survey respondent.

**Other:** any other tool used that is not Psiphon, Lantern, V2Ray, DNS services mentioned by the respondent. This includes VPNs that are infrequently mentioned like BiuBiu or Argo.

**Multiple VPNs:** more than 2 VPNs are reportedly being used by the survey respondent.

**Why heatmaps:** I wanted to find a way to visualize data for the open-ended questions, so I began with this methodology. I also discovered that heatmaps help visualize user behavior.

*Question: Please specify which VPN, DNS and/or other methods you use when accessing online video games and for which games.*

The gaming tool usage analysis across 523 users in the top 50 cities reveals a distinctly different pattern from web browsing preferences. "Other" methods lead the usage at 155 users (29.6%), followed closely by generic VPN solutions with 127 users (24.3%). Psiphon, while dominant in web browsing, ranks third in gaming with 99 users (18.9%).

The distribution shows similar patterns but with higher concentrations in the top 15 cities (415 users). "Other" methods account for 125 users (30.1%), Generic VPN for 99 users (23.9%), and Psiphon for 72 users (17.3%). The higher prevalence of Generic VPN solutions suggests that gamers may prioritize connection stability and speed over the specific features offered by specialized tools like Psiphon.

Notably, DNS Services show higher adoption in gaming (48 users, 9.2%) compared to web browsing, possibly due to their potential for lower latency. V2Ray maintains a minimal presence with only four users (0.8%), indicating it's not a preferred solution for gaming needs. The data suggests that gaming users tend to gravitate toward solutions that prioritize connection quality and consistency over other factors.

The heatmaps suggest that users employ different strategies for different online activities. Web browsing users favor specialized tools like the Psiphon, while gaming users opt for more general-purpose solutions and alternative methods.

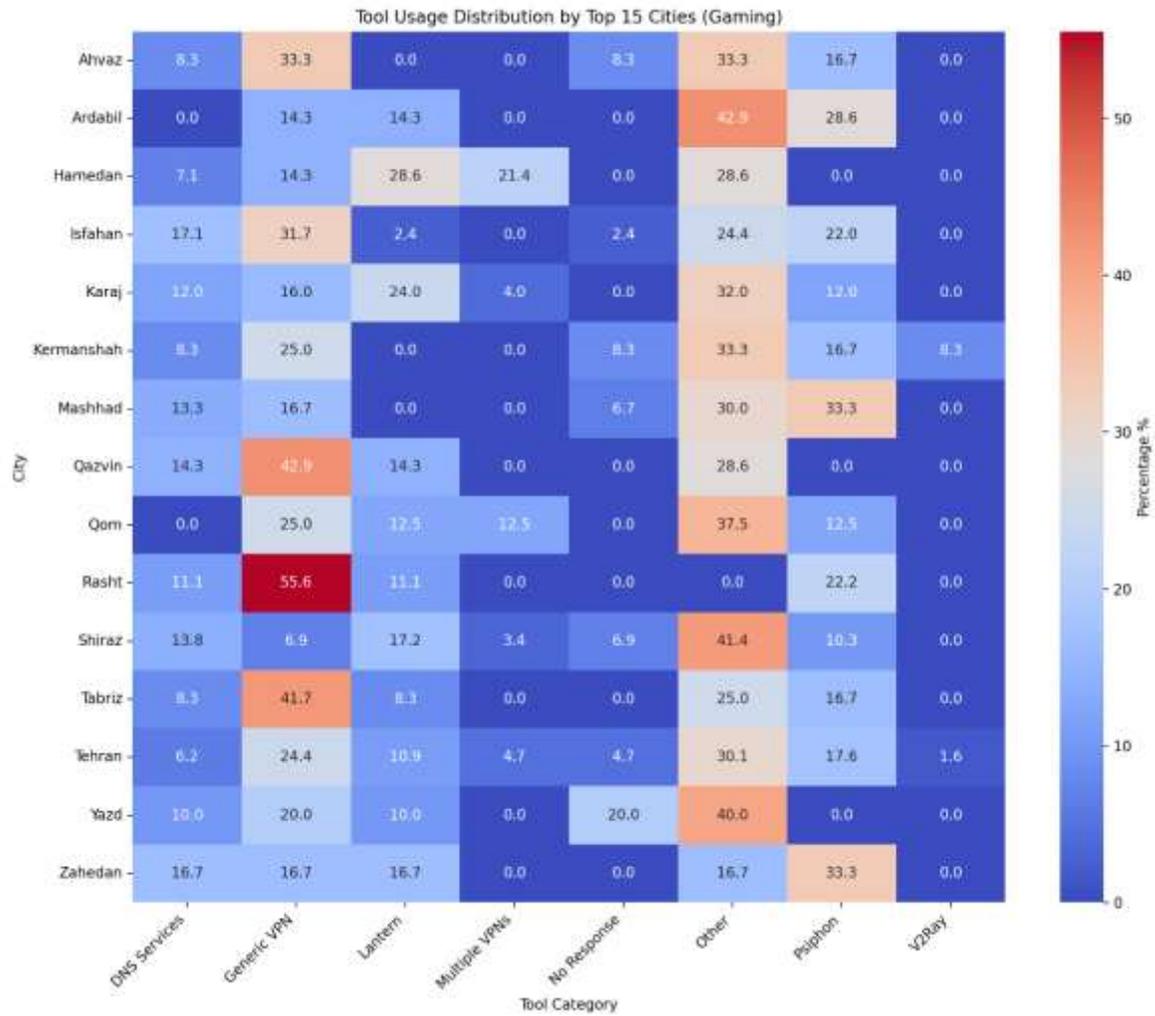

Figure 3. Heatmap represents the top 15 cities with 415 users out of the 660 respondents.

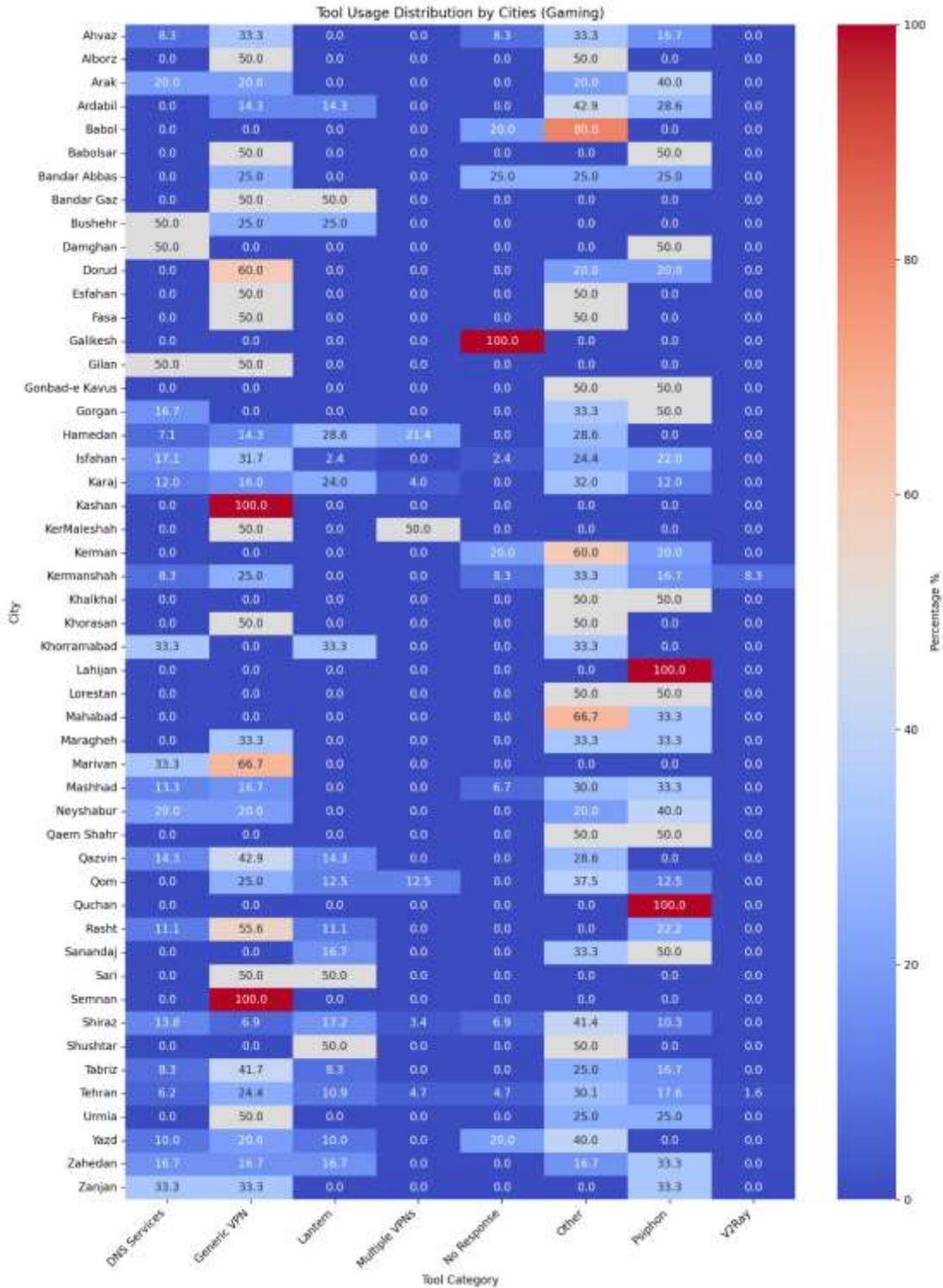

Figure 4. Heatmap represents the top 50 cities with 523 users out of the 660 respondents. 97 cities are left out due to only one user reporting in for each city.

The data covers 523 users across 50 cities, with a focus on the top 15 cities for detailed analysis. Tehran leads in gaming tool diversity with 8 unique tools used by 193 users. Shiraz follows with 7 tools and 29 users, while Isfahan has 6 tools and 41 users. Psiphon and generic VPNs dominate usage in these cities, reflecting their popularity for gaming.

The largest user base (213 users) is located in the northern part of Iran with balanced preferences for Psiphon (16.9%), general VPNs (25.8%), and "other" tools (29.1%). Meanwhile, the southern region has a smaller user base of 54 users with strong preferences for generic VPNs (33.3%) and "other" tools (33.3%).

The central region had a moderate user response rate of 66 users with notable preferences for generic VPNs (30.3%) and "other" tools (27.3%). The eastern region has high Psiphon usage (35.1%) among 37 users, indicating reliance on this tool for gaming, while the western region has the smallest reporting base of 33 users with a balanced distribution across tools.

DNS Services see higher adoption in gaming compared to web browsing. Tehran stands out with the highest diversity of tools, indicating a more tech-savvy user base or greater access to various solutions.

The top 5 cities account for 318 users (60.8% of the total), with Tehran alone contributing 193 users. Tool diversity is highest in Tehran and Isfahan. At the same time, tool preferences vary slightly across regions, with Psiphon leading in all areas. Users prioritize tools that offer reliability and ease of use for gaming, with Psiphon and Generic VPNs being the lead contender.

There could arguably be limited infrastructure in southern and eastern regions as the driver for higher reliance on Psiphon and "other" tools.

Improving internet infrastructure and access could reduce the reliance on bypass tools and enhance gaming experiences. The data likely does not fully capture rural areas or smaller cities, and self-reported tool usage could introduce bias. My goal is then to look at the web browsing VPN choices and compare these different data elements.

***Question: "If you use different filtering bypass tools when accessing the Internet for web browsing, please specify which one you use."***

Among the 523 total users across the top 50 Iranian cities, there is a clear preference for specific internet filtering bypass tools. Psiphon emerges as the dominant tool, used by 185 users (35.4%), followed by "other" methods by 125 users (23.9%). Lantern is the third most popular choice with 76 users (14.5%). V2Ray, while present, shows minimal adoption with only 3 users (0.6%).

The pattern remains consistent in the top 15 cities, which account for 415 users (79.3% of the total sample), but with slightly different proportions. Psiphon maintains its leading position with 137 users (33%), while "Other" methods and Lantern follow with 100 (24.1%) and 70 (16.9%) users, respectively. The data suggests that larger cities tend to have more diverse tool usage patterns, with users spread across multiple solutions rather than concentrating on a single tool.

DNS Services show limited adoption across both samples, indicating a strong preference for VPN-based solutions over DNS manipulation methods. Multiple VPN usage is more common in larger cities, suggesting greater technical sophistication or a need for redundancy in urban areas.

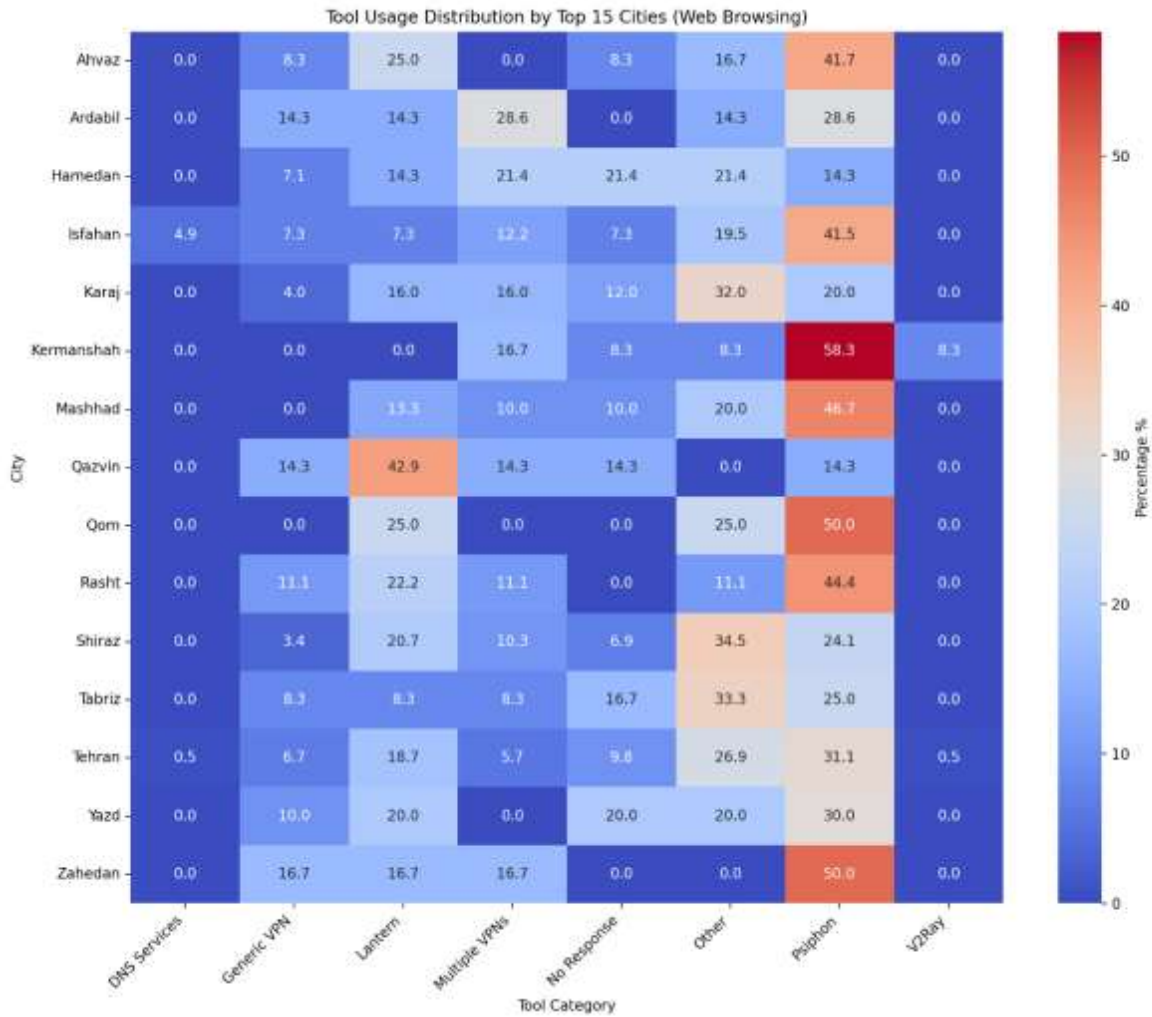

Figure 5. Heatmap represents the top 15 cities with 415 users out of the 660 respondents.

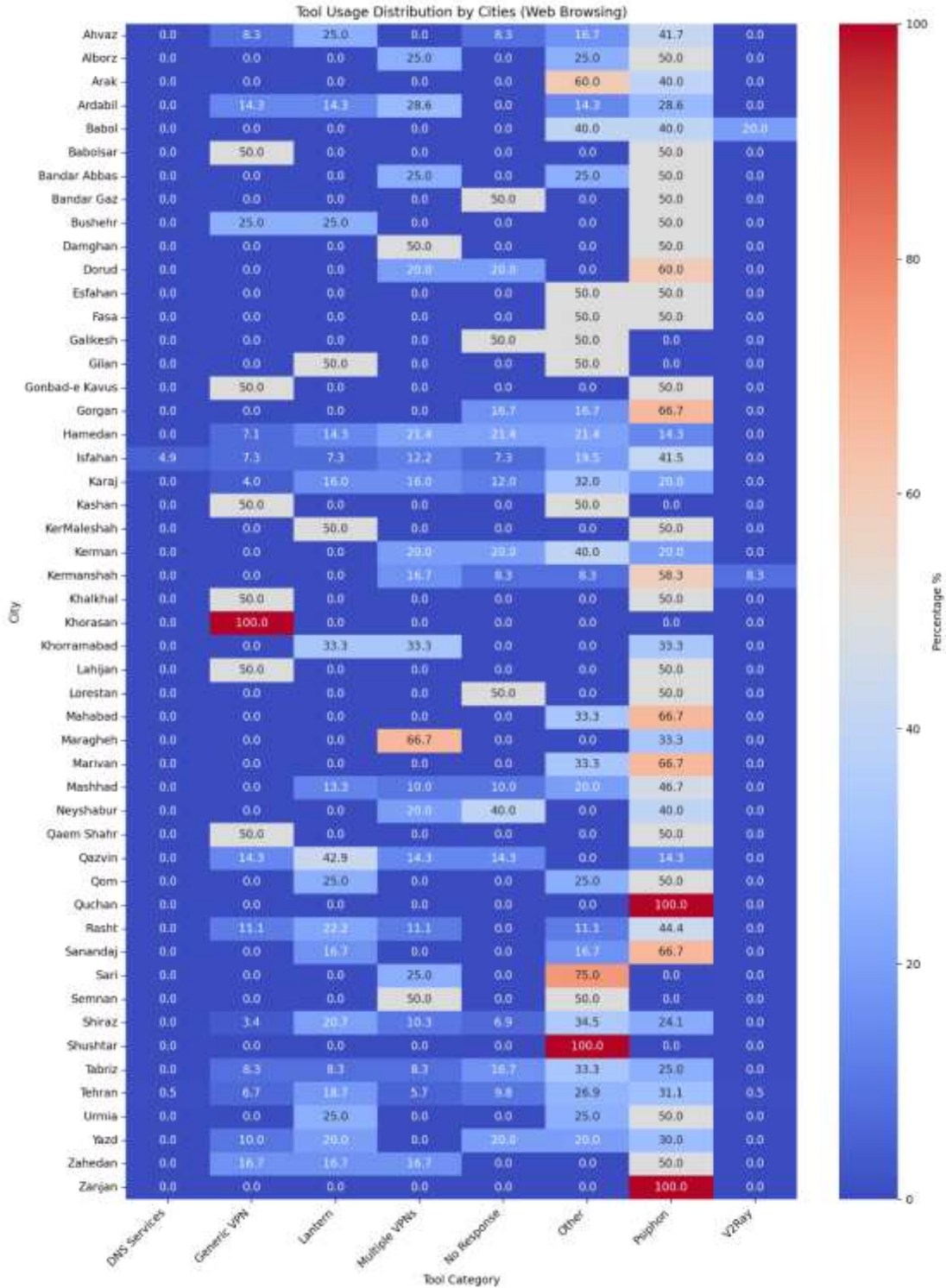

Figure 6. Heatmap represents the top 50 cities with 523 users out of the 660 respondents.

Tehran leads in tool diversity, with 8 unique tools used by 193 users. Isfahan follows with 7 tools and 41 users, while Shiraz has 6 tools and 29 users. Psiphon dominates usage in these cities, reflecting its popularity for web browsing.

The northern region had the largest user base (213 users) respond to the survey with balanced preferences for Psiphon (30.9%), "other" tools (27.7%), and Lantern (18.3%). Meanwhile, the southern region had a smaller user base (54 users) reporting in with strong preferences for Psiphon (39.4%) and "other" tools (24.2%).

The central region had a moderate user base (66 users) with notable preferences for Psiphon (39.4%) and "Other" tools (24.2%). The eastern region had a high Psiphon usage (45.9%) among 37 users, indicating reliance on this tool for web browsing. The western region had the smallest user base (33 users) with a balanced tool distribution.

Overall, Psiphon is the dominant tool across all regions, with its highest usage in the Eastern region. DNS Services see minimal adoption for web browsing. Tehran is the most unique case because it has a high level of tool diversity, reflecting a more tech-savvy user base and greater access to various solutions. The top 5 cities account for 318 users (60.8%), with Tehran alone contributing 193 users.

Tool diversity is highest in Tehran and Isfahan. Tool preferences are consistent across regions, with Psiphon leading in all areas. "Other" tools show significant usage in the North and Central regions. Users prioritize tools that offer reliability and ease of use for web browsing, with Psiphon being the clear favorite. As suggested in the first question's conclusion on game tool use, the same findings can be suggested here: limited infrastructure in Southern and eastern regions may drive higher reliance on Psiphon and "other" tools. Overall, my hopeful goal is to see improvements in internet infrastructure to enhance both web browsing and gameplay experiences.

**Hierarchical Regression Analysis: Closed-Ended Questions**

The models below are based on the survey questions corresponding to the following dependent variables:

1. "I feel confident about accessing games." This measures confidence in accessing games.
2. "I am confident that I can access games safely." This measures confidence in safe access.
3. "I am confident that I can maintain privacy while playing." This measures confidence in maintaining privacy.

   <u>Definitions</u>
- Y is the dependent variable, like "I feel confident about accessing games."
- $\beta_0$ is the intercept.
- $\beta_1, \beta_2, \beta_3 \beta_1, \beta_2, \beta_3$ are the coefficients for the predictors.
- $\epsilon$ is the error term.

Demographics Formula

$Y = \beta_0 + \beta_1(\text{Age Group}) + \beta_2(\text{Education Level}) + \epsilon$

Adding Confidence to Formula

$Y = \beta_0 + \beta_1(\text{Age Group}) + \beta_2(\text{Education Level}) + \beta_3(\text{Confidence in Using Bypass Tools}) + \epsilon$

Findings
- Dependent Variable: "I feel confident about accessing games"
  - R-squared: 0.5998 (59.98% of the variance explained).
  - Significant Predictors: Age group 40+ ($p<0.001 p<0.001$) negatively impacts confidence.
  - Education levels were not significant predictors.
- Dependent Variable: "I am confident that I can access games safely"
  - R-squared: 0.5132 (51.32% of the variance explained).
  - Significant Predictors: None of the demographic variables were significant, indicating confidence in bypass tools may play a larger role.
- Dependent Variable: "I am confident that I can maintain privacy while playing"
  - R-squared: 0.3958 (39.58% of the variance explained).
  - Significant Predictors: Age group 40+ ($p=0.067 p=0.067$) showed a marginally significant negative impact.

Interpretation
- Age Group 40+ consistently negatively impacts confidence across models, suggesting older individuals may feel less confident in accessing or maintaining privacy while playing games.
- Education Level does not significantly predict confidence, indicating that other factors (e.g., experience with bypass tools) may be more influential.
- Confidence in Using Bypass Tools likely explains a significant portion of the variance, as seen in the high R-squared values.

Data Table & Additional Insights

| Dependent Variable | R-squared | F-statistic | p-value | Age 20-30 Coef (p-value) | Age 30-40 Coef (p-value) | Age 40+ Coef (p-value) |
|---|---|---|---|---|---|---|
| Accessing Games Confidence | 0.5998 | 74.4803 | 0 | -0.118 (0.299) | -0.134 (0.157) | -0.317 (0.000) |
| Safe Access Confidence | 0.5132 | 52.3921 | 0 | -0.026 (0.838) | 0.105 (0.329) | -0.009 (0.928) |

| Privacy Maintenance Confidence | 0.3958 | 32.5555 | 0 | -0.066 (0.653) | -0.094 (0.448) | -0.211 (0.067) |

Figure 7. Data table based on confidence to accessing games

1. The model explaining confidence in accessing games has the highest R-squared (0.5998). Privacy maintenance confidence has the lowest R-squared (0.3958)
2. The age 40+ group shows the strongest effects across models. The negative effect is most pronounced for accessing games confidence (-0.317, $p < 0.001$). Marginally significant effect on privacy maintenance (-0.211, $p = 0.067$)
3. All models are statistically significant ($p < 0.001$) based on F-statistics. Younger age groups (20-30 and 30-40) show no significant effects

### 3.3.3 Network Analysis

RIPE Atlas Data

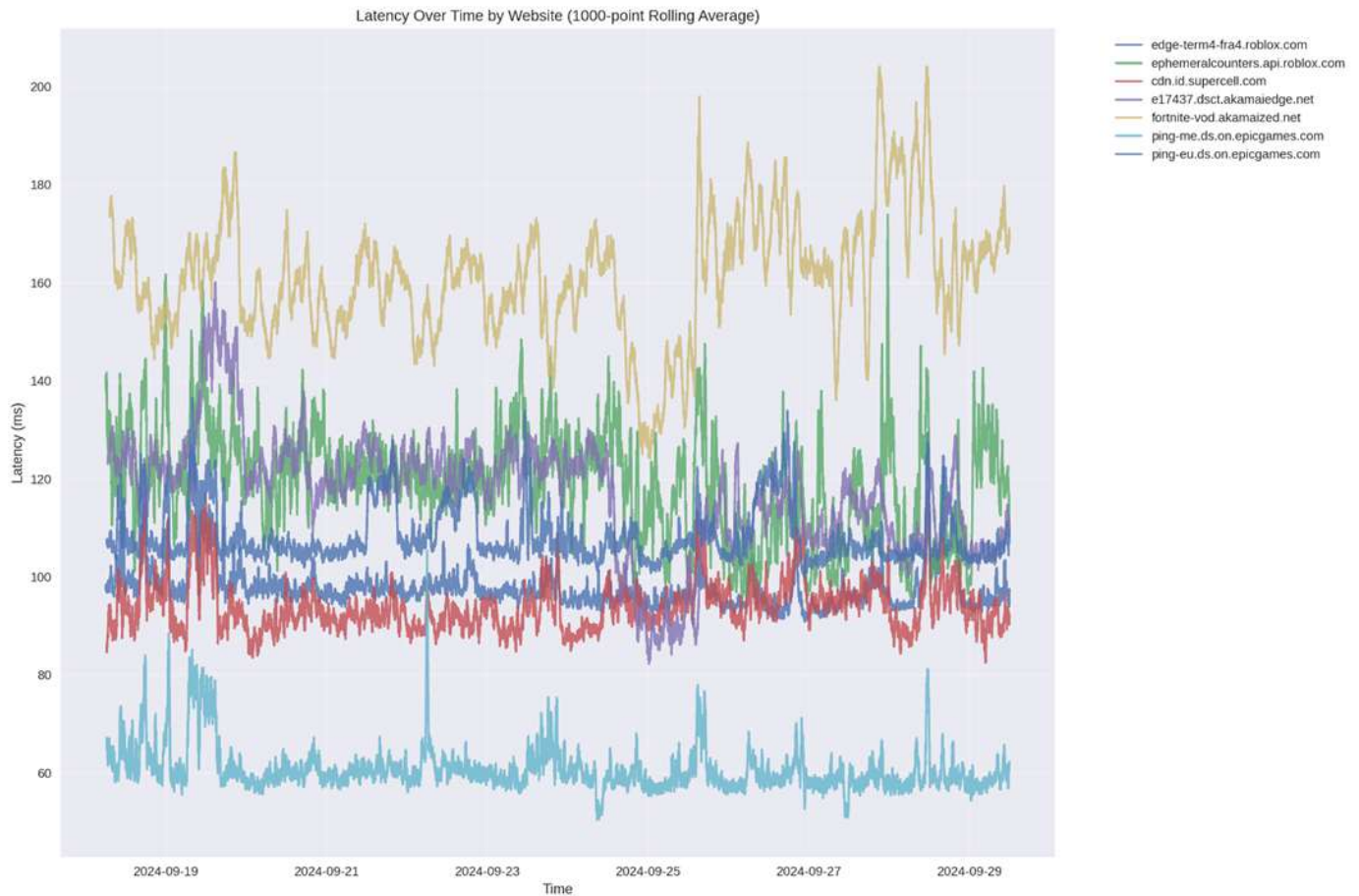

Figure 8. RIPE Atlas collected across 7 gaming websites

1. Epic Games Middle East (ping-me.ds.on.epicgames.com):
    a. Best performing server with ~93% of measurements in the playable range (under 100ms)
    b. Excellent for competitive gaming with 48% of connections under 50ms
    c. Suitable for fast-paced games like Fortnite
2. Game Servers (Roblox, Supercell):
    a. Generally good performance (60-65% under 100ms)
    b. Roblox edge servers show consistent performance
    c. Supercell CDN maintains good stability for mobile gaming
3. Content Delivery (Fortnite VOD, Akamai):
    a. Higher latencies but less critical for gameplay
    b. Fortnite VOD shows 48.79% measurements above 150ms
    c. Primarily impacts game downloads and updates, not actual gameplay

| Website | Avg. Latency | Stability (std) | Excellent (<20ms) | Very Good (20-50ms) | Good (50-100ms) | Playable (100-150ms) | Issues (>150 ms) | Stability Score |
|---|---|---|---|---|---|---|---|---|
| ping-me.ds.on.epicgames.com | 60.85 | 55.87 | 6.15 | 42.11 | 44.88 | 1.78 | 5.08 | 58.858 |
| cdn.id.supercell.com | 94 | 54.02 | 0 | 14.09 | 47.09 | 33.4 | 5.42 | 78.008 |
| edge-term4-fra4.roblox.com | 98.14 | 43.17 | 0 | 0 | 63.63 | 34.5 | 1.87 | 76.152 |
| ping-eu.ds.on.epicgames.com | 108.52 | 47.69 | 2.06 | 0 | 41.89 | 51.37 | 4.69 | 84.188 |
| e17437.dsct.akamaiedge.net | 117.02 | 77.56 | 0 | 18.97 | 24.2 | 31.73 | 25.1 | 101.236 |
| ephemeralcounters.api.roblox.com | 119.47 | 72.35 | 0 | 0 | 52.09 | 35.96 | 11.95 | 100.622 |
| fortnite-vod.akamaized.net | 161.4 | 85.65 | 0 | 3.31 | 19.41 | 28.49 | 48.79 | 131.1 |

Figure 9. Websites measure for latency and gaming capacity.

The stability scores show which services provide the most consistent gaming experience, with lower scores indicating better performance. This is crucial because:

- Consistent latency (even if slightly higher) is often better for gaming than lower but unstable latency
- Players can adapt to consistent latency, but unpredictable spikes disrupt gameplay

- The ME and EU Epic Games servers show the best stability, ideal for competitive gaming

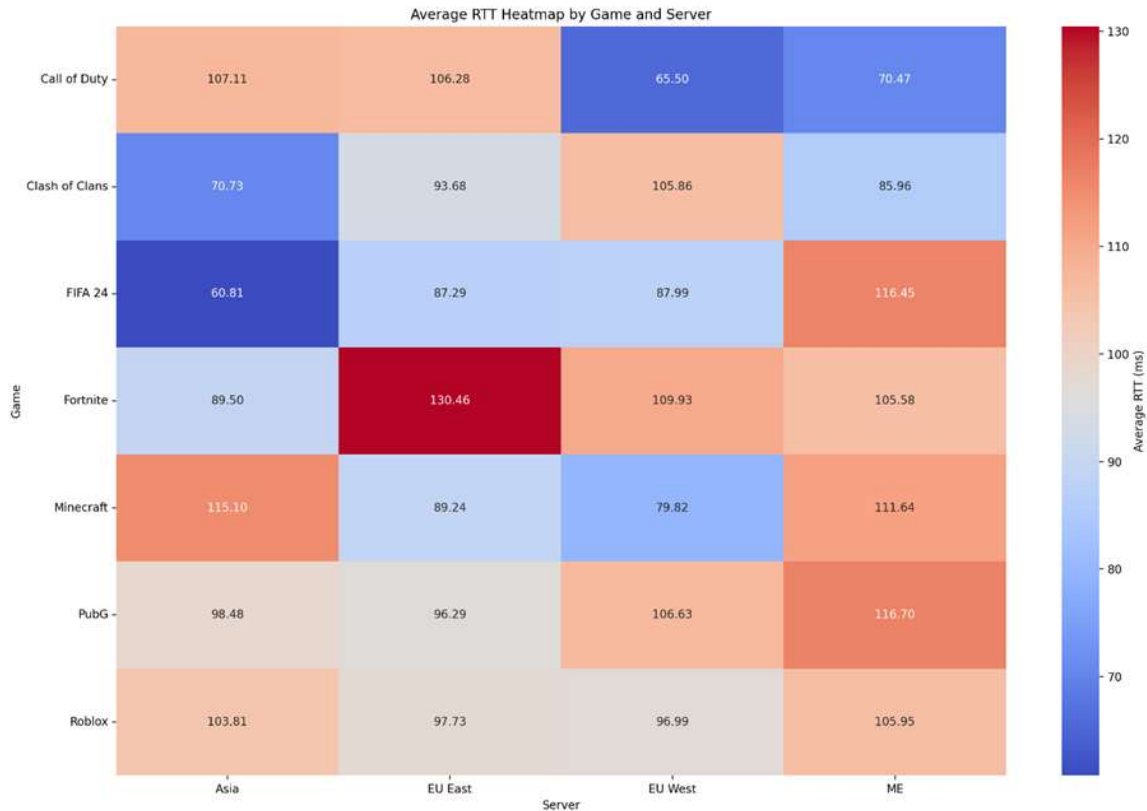

Figure 10. Average RTT heatmap by game and server.

The analysis of gaming performance metrics, mainly focusing on Round Trip Time (RTT), provides valuable insights into internet access in Iran. The Middle East (ME) server, geographically relevant to Iran, exhibits an average RTT of 105.17 ms, with a standard deviation of 12.36 ms. This RTT is relatively higher than that of European servers, such as EU East and EU West, which have average RTTs of 97.51 ms and 98.95 ms, respectively. These metrics show the challenges Iranian users face in accessing low-latency internet services. High RTT values can significantly impact real-time applications like online gaming, video conferencing, and other latency-sensitive activities. To collect this data, Ainita Project conducted 2,185 measurements and 215 probes across the ME server, indicating a substantial amount of data collected to assess performance.

The importance of internet access in Iran extends beyond gaming. Reliable and low-latency internet is crucial for education, remote work, and communication, especially in a region where digital connectivity can bridge gaps in infrastructure and access to global resources. The higher

RTT values observed for the ME server emphasize the need for investments in internet infrastructure, including better routing, increased bandwidth, and reduced packet loss. In conclusion, the data-driven analysis of RTT and related metrics sheds light on Iran's internet access state. Addressing these challenges is essential for empowering Iranian users and ensuring equitable access to the digital world.

## 4. Conclusion

Exploring Iranian gamers' resilience against pervasive state-imposed digital restrictions has revealed a multifaceted landscape of technical ingenuity and grassroots solidarity. This report has delved into the layered effects of digital authoritarianism, illustrating that resistance in such environments is never one-dimensional; it is a convergence of sophisticated technical know-how and dynamic social collaboration. Iranian gamers, trapped in a web of surveillance and censorship, not only strive to stay connected but also to dismantle the imposed boundaries that restrict their digital lives. This comprehensive investigation has integrated qualitative narratives, quantitative survey data, and technical network analysis to paint a vivid picture of how a community under duress redefines its relationship with technology.

At the heart of this study lies a powerful demonstration of human adaptability. The qualitative interviews capture gamers' daily struggles and creative workarounds, highlighting practices ranging from the innovative use of VPNs and encryption tools to the formation of supportive online networks that enable shared knowledge. These narratives showcase a blend of frustration, resilience, and optimism—a reminder that every technical workaround is anchored in a deeply personal experience of defiance. Yet, these personal accounts are more than isolated instances; they are representative of a broader, communal effort to reclaim digital space from the grips of repression. The survey data reinforces this perspective, outlining patterns of behavior and trust that underscore the community's collective approach. Differences in demographic trends and varying degrees of technical proficiency across age groups illustrate how digital resistance is tailored to diverse needs and challenges.

Moreover, the technical network analysis provides an objective lens on the practical implications of censorship. It indicates that, while Iranian gamers face significant latency and stability issues compared to peers accessing alternative networks, they have adeptly adapted their technical strategies to offset these challenges. This part of the study underscores that innovation is not solely a reaction to repression but a driver of change in how digital infrastructures are used and understood. The tension between state control and community-based innovation is palpable—it is a dynamic interplay where technological limitations spur creative problem-solving and localized adaptation.

In merging these threads, the report reveals that digital resilience is more about forging robust social networks than employing advanced technology. The convergence of personal experiences, statistically significant survey outcomes, and challenging technical measurements provide concrete evidence of a sophisticated resistance mechanism. This integrated approach goes beyond merely outlining the struggles; it highlights the ingenuity and perseverance of individuals continuously striving for digital freedom under oppressive conditions. The qualitative, quantitative, or technical methodologies adopted in this study serve as a holistic blueprint for future research into digital resistance. They suggest that engaging with such multiplicity is necessary for a nuanced understanding of digital authoritarian regimes and is pivotal in shaping policy recommendations that support an open and accessible internet.

In essence, the resilience documented in this report is not a fleeting act of defiance against a transient challenge. Instead, it is an ongoing, systemic response to a rigorously enforced digital landscape that continuously evolves. The findings remind us that, amid stringent control, there is an undercurrent of innovative and profound communal resistance. As state mechanisms of control intensify, the creative strategies employed by these gamers offer a beacon of hope and a challenge to the status quo. They compel policymakers, technology developers, and global digital rights advocates to rethink existing approaches and to support the structures that nurture such resilient communities.

Thus, the report serves as a detailed academic inquiry and a pragmatic call to action. It underscores the need for deliberate policy measures, robust infrastructure support, and international cooperation to safeguard digital freedom. The implications of this work extend beyond the immediate context of Iranian gamers; they resonate with anyone who values the power of community and technology to resist oppressive systems. As we continue to witness the evolution of digital landscapes globally, the lessons drawn from this study will remain pertinent, reminding us that resilience is a potent force that not only survives censorship but transforms it into an impetus for collective innovation and future progress.

## 5. Ongoing and Future Work

In 2020, Digital Iran began as a collaborative project to understand the dynamics between the Iranian video game industry and video game content about government propaganda (Cohoon 2022). With the help of stakeholders during the third cycle of the project funded by OTF, Digital Iran

Building on the current findings, future work will deepen the exploration of Iranian gamers' sociotechnical resilience and extend the analysis into two key areas: gendered differences in digital navigation strategies and the transformative role of artificial intelligence in digital media analysis.

Preliminary findings indicate that social positionality and gender significantly shape how Iranian gamers experience online censorship and deploy circumvention techniques. Ongoing work will involve a more granular analysis of gender-based resilience strategies. This includes expanding survey analysis, conducting targeted interviews, and integrating ethnographic observations to understand the interplay between cultural norms and digital expression. Insights gained here will refine our theoretical framework on sociotechnical resilience in the context of a repressive digital landscape. Given the growing threat of malicious AI applications in disinformation and digital authoritarianism, we will design experiments to assess how such tools manipulate online content and censor digital narratives. By integrating network analysis with machine learning models, we aim to identify vulnerabilities in Iran's internet infrastructure that malicious AI might exploit.

In addition to tackling gendered dynamics, Digital Iran is also seeking funding in other domains to expand its initiative by working with other academics to enhance data extraction and annotation of audiovisual and textual media from video games. This effort will support the creation of a comprehensive annotated dataset to capture cultural nuances and resistance tactics embedded in digital media.

Drawing upon the insights obtained through AI-driven analysis, we will delineate policy recommendations for mitigating the misuse of AI in shaping public opinion and enforcing censorship. Equally, our work will inform the design of next-generation circumvention tools that harness AI's analytical power while safeguarding against its potential misuse.

Future efforts will also deepen collaborations with renowned organizations in digital humanities, cybersecurity, and policy research. As we look to additional funding and partnerships, we look to the same entities that assisted the project during its tenure at OTF, which will be pursued to triangulate technical analysis with real-world impacts. This interdisciplinary approach will help refine the methodologies and ensure that the work remains at the cutting edge of research on digital authoritarianism and AI's mitigating potential.

Finally, we plan to extend the temporal scope of our study through longitudinal tracking of digital media trends and censorship practices over time. This extended observation, paired with transnational data collection—including analysis of the Iranian diaspora's role in combating online disinformation—will enrich our understanding of how digital communities reconfigure resistance strategies in the face of evolving technological challenges.

Digital Iran aims to document interlinked aspects of Iranian digital resistance and contribute actionable insights through this proposed future work. It is currently pioneering a new framework for analyzing and countering digital authoritarianism in complex cultural landscapes by harnessing AI's capabilities, both its promise and pitfalls.

## 6. Supplementary Materials

Survey Questions

1. کشور

2. شهر
3. سن
4. جنسیت
5. تحصیلات
6. دین
7. اقلیت قومی/مذهبی
8. وضعیت اشتغال
9. به صورت عمومی چقدر از فیلترشکن استفاده می‌کنید؟

من هرگز چیزی در مورد فیلترشکن نشنیدم
من در مورد فیلترشکن شنیدم ولی از آن استفاده نکردم
من چندین بار از فیلترشکن استفاده کردم
من گاهی از فیلترشکن استفاده می‌کنم
من معمولا از فیلترشکن استفاده می‌کنم
من همیشه از این ابزارها استفاده می‌کنم

10. چه ارایه دهنده‌های اینترنتی در محدوده شما وجود دارند؟ و شما از کدام استفاده می‌کنید؟

11. اگر هنگام دسترسی به اینترنت برای وب گردی از ابزارهای دور زدن مختلفی استفاده می‌کنید، لطفاً مشخص کنید که از کدام یک استفاده می‌کنید.

12.
به طور کلی، چقدر از ابزارهای دور زدن فیلترینگ برای دسترسی به بازی‌های ویدیویی آنلاین استفاده

من در مورد فیلترشکن شنیدم ولی از آن استفاده نکردم
من چندین بار از فیلترشکن استفاده کردم
من گاهی از فیلترشکن استفاده می‌کنم
من معمولا از فیلترشکن استفاده می‌کنم
من همیشه از این ابزارها استفاده می‌کنم

13. لطفاً مشخص کنید که از کدام DNS، VPN و/یا روش‌های دیگر هنگام دسترسی به بازی‌های ویدیویی آنلاین و برای کدام بازی‌ها استفاده می‌کنید.

14. لطفاً سطح اعتماد خود را با استفاده از ابزارهای دور زدن هنگام بازی آنلاین نشان دهید.
(کاملا مخالف / مخالف / نه موافق و نه مخالف / موافق / کاملا موافق)
من با استفاده از ابزارهای دور زدن برای انجام بازی‌های آنلاین احساس اطمینان می کنم.
از دسترسی به بازی‌هایی که می خواهم به صورت آنلاین بازی کنم احساس اطمینان می کنم.
من مطمئن هستم که می توانم به طور ایمن به بازی های آنلاین دسترسی داشته باشم
من مطمئن هستم که می توانم حریم خصوصی خود را در حین بازی آنلاین حفظ کنم.

15. لطفاً سطح پینگی را که هنگام استفاده از ابزارهای دور زدن هنگام بازی تجربه می کنید، مشخص کنید. کدام بازی ها و چه سرورهایی؟ و آیا این تفاوت دارد؟

16.

لطفاً به ما بگویید که چقدر با جملات زیر در مورد برنامه هفتم توسعه ایران موافق یا مخالف هستید

(کاملا مخالف / مخالف / نه موافق و نه مخالف / موافق / کاملا موافق)

برنامه هفتم توسعه ایران بر دسترسی من به اطلاعات آنلاین تأثیر می گذارد.
برنامه توسعه هفتم ایران بر دسترسی من به بازی آنلاین تأثیر می گذارد
برنامه هفتم توسعه ایران نگرانی بزرگی است.
برنامه هفتم توسعه ایران نباید دسترسی به اینترنت را طبقه بندی کند.

1. Country
2. City
3. Age
4. Sex
5. Education
6. Religion
7. Ethnicity
8. Employment
9. In general, how often do you use circumvention tools to access the web?

    -I have never heard of circumvention tools before today.
    -I have heard of circumventions tools, but I have never used them to access the internet.
    -I have used circumvention tools a few times.
    -I use circumvention tools occasionally.
    -I use these tools regularly.
    -I use these tools all the time.

10. What are the internet service providers in your area? And which internet service provider do you use?

11. If you use different circumvention tools when accessing the internet for web browsing, please indicate which ones you use.

12. In general, how often do you use circumvention tools to access online video games?

    -I have never heard of circumvention tools before today.
    -I have heard of circumventions tools, but I have never used them to access the internet.
    -I have used circumvention tools a few times.
    -I use circumvention tools occasionally.
    -I use these tools regularly.
    -I use these tools all the time.

13. Please indicate which VPNs, DNSs, and/or other methods you frequently use when accessing online video games, and for which games.

14. Please indicate your confidence level using circumvention tools when playing games online.

    (Strongly disagree/Disagree/Neither agree nor disagree/Agree/Strongly agree)

    -I feel confident using circumvention tools to play games online.
    -I feel confident accessing the games I want to play online.
    -I feel confident that I can securely access games online.
    -I feel confident I can maintain my privacy while playing online games.

15. Please indicate the ping level you experience when using circumvention tools while playing games. Which games and what servers? And does this vary?

16. Please tell us how much you agree or disagree with the following statements about Iran's seventh development plan.

    (Strongly disagree/Disagree/Neither agree nor disagree/Agree/Strongly agree)
    -The Iranian seventh development plan impacts my access to information online.
    -The Iranian seventh development plan impacts my access to gaming online.

    -The Iranian seventh development plan is a great concern.

    -The Iranian seventh development plan should not tier internet access.

## 7. Acknowledgments

I would like to express immense gratitude to the Open Tech Fund for their generous funding. This research would not be possible without their support. The Information Controls Fellowship has afforded me the tremendous opportunity to meet new researchers, discover new ways of thinking, and present at the Global Gathering among internet freedom's finest. I am also grateful to the host organization, the Miaan Group, who helped me throughout the research process. Special thanks to Amir Rashidi for your kindness and support.

Amini's death in custody needs urgent global action. https://www.amnesty.org/en/latest/news/2022/09/iran-deadly-crackdown-on-protests-against-mahsa-aminis-death-in-custody-needs-urgent-global-action/

Article 19 (2022, September 09). Iran: Cyberspace authorities 'silently' usher in draconian internet bill. https://www.article19.org/resources/iran-draconian-internet-bill/

Article 19. (2024, July 23). Tightening the Net: Iran's New phase of digital repression. https://www.article19.org/resources/tightening-the-net-irans-new-phase-of-digital-repression/.

Aryan, S., Aryan, H., & Halderman, J.A. (2013). Internet Censorship in Iran: A First Look. *FOCI 13*. https://www.usenix.org/system/files/conference/foci13/foci13-aryan.pdf

Bacovsky, P. (2020). Gaming alone: Videogaming and sociopolitical attitudes. *New Media & Society*, *23*(5), 1133–1156. https://doi.org/10.1177/1461444820910418

Cohoon, M. (2022). Digital Borderlands: Soft War as Discourse in Iranian Video Games. *IDEAH, 3*(2). https://doi.org/10.21428/f1f23564.73008090

Cohoon, M. (2022). Information controls in Iranian cyberspace: a soft war strategy. In Case Analysis. https://www.dohainstitute.org/en/Lists/ACRPS-PDFDocumentLibrary/information-controls-in-iranian-cyberspace-a-soft-war-strategy.pdf

Cohoon, M. (2022, December 1). Presented at the Middle East Studies Association Conference. https://melindacohoon.com/presented-at-the-middle-east-studies-association-conference/

Corera, G. (2021, February 8). Iran 'hides spyware in wallpaper, restaurant and game apps." BBC. https://www.bbc.com/news/technology-55977537

Dadbazar. (2024, February 1) Resolution of the Supreme Cyberspace Council on Combating Filter Violators. https://www.ekhtebar.ir/%d9%85%d8%b5%d9%88%d8%a8%d9%87-%d8%b4%d9%88%d8%b1%d8%a7%db%8c-%d8%b9%d8%a7%d9%84%db%8c-%d9%81%d8%b6%d8%a7%db%8c-%d9%85%d8%ac%d8%a7%d8%b2%db%8c-%d8%af%d8%b1%d8%ae%d8%b5%d9%88%d8%b5-%d9%85%d9%82%d8%a7/.

Dehshiri, A. (2024, May 17). #The use of VPNs is prohibited, but not criminalized. Filterwatch. https://filter.watch/english/2024/03/04/the-use-of-vpns-is-prohibited-but-not-criminalized/

Egherman, T. (2025, March 19). Iran's Digital Control: the evolution of censorship and surveillance amidst the 'Women, Life, Freedom' movement. Miaan Group. https://miaan.org/irans-digital-control-the-evolution-of-censorship-and-surveillance-amidst-the-women-life-freedom-movement/

Farda-ye Eghtesad. (2023, October 24). The government is training 500,0000 people for cyberspace +video
https://www.fardayeeghtesad.com/news/31420/%D8%AF%D9%88%D9%84%D8%AA-